# Memory effects in Amorphous Solids below 20 mK

D. Rosenberg, P. Nalbach, and D. D. Osheroff
*Department of Physics, Stanford University, Stanford, CA 94305, USA*
(Dated: January 11, 2003)

At temperatures below 1 K, the capacitance of a glass sample changes due to the application of a DC field in accordance with A. Burins dipole gap theory. However, we now report that below 20 mK, during the first sweep cycle of the DC electric field the capacitance is smaller by about $10^{-5}$ compared to any subsequent sweep. Despite this overall shift the field dependence follows the dipole gap predictions. In a subsequent sweep to higher DC fields the dielectric constant drops by about $10^{-5}$ as soon as the applied field is higher than any field previously applied. A picture involving the dynamics of resonant pairs provides a qualitative description of this behavior.

The properties of amorphous solids, which behave in a fundamentally different way from crystalline solids at temperatures below 1 K [1], are well described by the two level system (TLS) model, first proposed in 1972 [2, 3]. This model assumes the existence of a large distribution of tunneling systems which can interact with strain and electric fields. Each tunneling system is characterized by some energy asymmetry $\Delta$ and a tunneling matrix element $\Delta_0$, which are distributed according to $P = P_0/\Delta_0$, where $P_0$ is a constant.

In the early 90's, Salvino et al [4] found that the AC capacitance of amorphous samples below 1 K is affected by DC fields. When a large DC electric field is applied to a glass sample, the capacitance is observed to jump up with the application of the field and then decay logarithmically in time while the field is applied. Sweeping the DC field slowly traces out a minimum at zero field. The DC field shifts the asymmetry energies of the TLS by $\vec{p} \cdot \vec{F}$, where $\vec{p}$ is the dipole moment of the TLS and $\vec{F}$ is the applied field. Carruzzo and Yu [5] show that the observed frequency dependence is inconsistent with a model of non-interacting TLS and suggest that interactions between the TLS lead to a hole in the density of states. Building on this idea, in 1995 A. Burin proposed a theory which considers a weak interaction between the two level systems and shows that a gap in the density of states, known as the dipole gap, forms at zero local field [6]. In this picture, the jump and subsequent decay in the dielectric constant are due to the destruction and reformation of the gap. Burins theory provides both a qualitative and quantitative description of the experiments.

We have now investigated the capacitance during the first few sweep cycles after cooling down the sample from above 1 K. Below 20 mK the first sweep through any value of the DC field results in a capacitance which is smaller than in subsequent sweeps by approximately one part in $10^5$. This leads to the sample essentially "remembering" the applied field values until its temperature is raised to 1 K.

All samples were immersed in $^3$He and the leads were heat sunk using the double pressed powder heat exchanger first introduced by Rogge et. al. [7]. Details of the immersion cell can be found in [8]. The experimental cell was bolted directly to a copper demagnetization stage attached to an Oxford 400 dilution refrigerator, and cooling of the $^3$He volume was accomplished using a pressed silver powder heat exchanger. The temperature was obtained from a $^3$He melting curve thermometer calibrated against the superfluid $^3$He transitions.

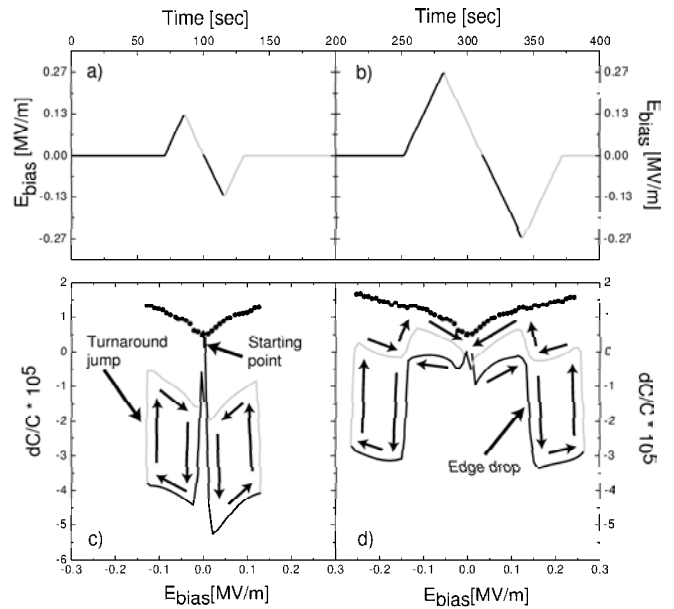

FIG. 1: Smoothed data showing the initial and edge drops and turnaround jump. The heavy curves at the top of c) and d) represent the capacitance seen after several sweeps.

The capacitance was measured using a standard analog bridge at frequencies between 1 and 10 kHz, and the excitation voltages used were 1 kV/m to 10 kV/m, which correspond to a temperature perturbation of 0.24 to 2.4 mK for a dipole moment of 1 Debye. These excitation voltages, while they are not so high that they exceed the temperature, place the measurements in the non-linear regime for equilibrium data as discussed in [9]. However, the effects described here do not depend strongly on the excitation voltage used.

Fig. 1 shows a typical experiment in which the elec-tric field across a sample is swept in a triangle wave with an amplitude $F_1$ after being cooled down from 1 K. A large decrease in capacitance is observed as the sweep begins. As the field continues to ramp up, the capacitance rises with the field as expected from the dipole gap theory. When the sweep reaches $F_1$ and reverses direction, the capacitance is observed to rise quickly (the "turnaround jump") , and as the field is decreased further the capacitance drops again according to the dipole gap theory, tracing out a line offset by a constant from the data taken while the field was ramped up.

If the sweep is repeated, the drop seen at the start of the sweep is much smaller, and it completely disappears after a few passes of the field, giving the standard dipole gap result. The heavy upper curves in Fig. 1c,d are data which were obtained after the field was swept back and forth several times so that only the dipole gap behavior was present.

The above behavior seems to indicate that there is something different in the glass the first time the electric field reaches a given value. A striking demonstration of this is seen if a second sweep with a larger amplitude $F_2$ but the same sweep rate is performed. When the magnitude of the electric field is less than $F_1$, the regular dipole gap effect is observed. As soon as the amplitude passes $F_1$, the capacitance quickly decreases, taking on the value it obtained just below $F_1$ during the first sweep. This sudden decrease, which we will call the "edge drop", marks the DC field amplitude for the first set of sweeps, so the system has essentially retained a memory for the earlier sweeps. As the sweep reaches the maximum voltage and reverses direction, an increase in capacitance is again observed. Although it occurs at a different voltage from the turnaround jump, it is the same size as for the first sweep. If a third sweep with an even larger amplitude is now performed, the same memory effect occurs.

The memory of the system is not permanent, however, and if one waits a very long time the original behavior will return. Experimentally, after days we still could not completely regain the behavior of the very first sweep. Instead we had to warm the sample up above 1 K to erase all memory.

To quantify how the system "forgets", we can perform two sweeps with different peak fields and define as the waiting time the time which elapses between reaching the maximum voltage during the first sweep and passing that voltage on the second sweep. If the system retains no memory of the first sweep, then no drop should occur when the second sweep passes $F_1$. Fig. 2 shows the dependence of the memory effect on waiting time for data taken at 5 mK. Each point represents a pair of sweeps with a given waiting time, and the sample was warmed up to 1 K between each pair of sweeps to erase the memory. As we expect, the effect is smaller for longer waiting times, and the data for waiting times greater than 1 minute roughly follow a logarithm as a function of waiting time.

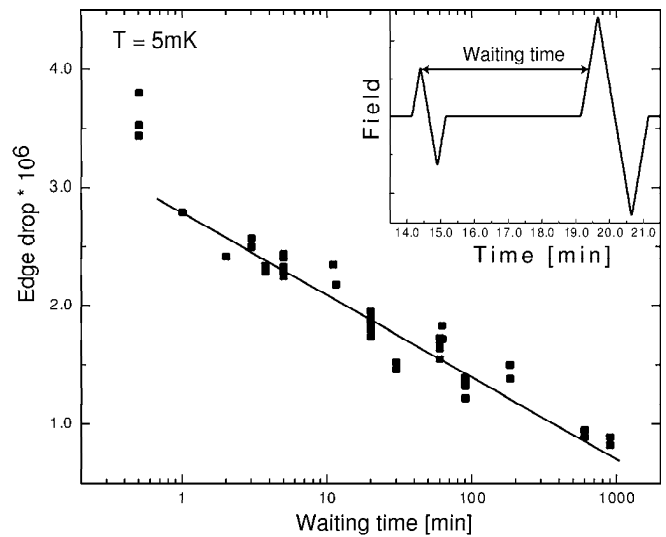

FIG. 2: The edge drop becomes smaller with increasing waiting time, defined in the upper right hand corner inset.

If the field is ramped up and then held at some value, the capacitance is observed to increase quickly as soon as the field stops changing. After the fast increase, a logarithmic decay which is consistent with the dipole gap theory is seen, as shown in the inset to Fig. 3. The main part of Fig. 3 shows the capacitance of a sample after the field has been ramped up and then is held at some value, with the dipole gap behavior subtracted out. The data shown were taken at 5 mK.

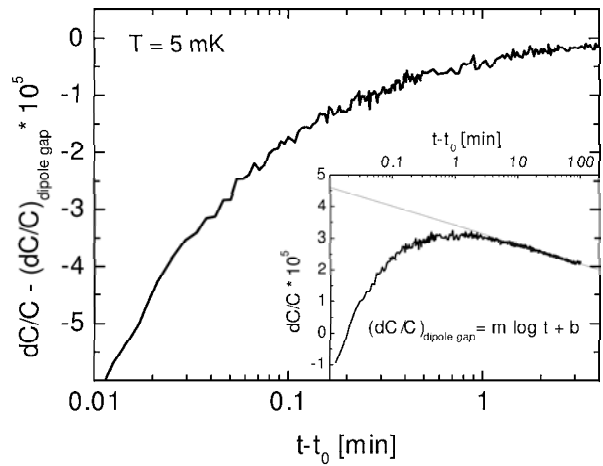

FIG. 3: When a DC field is ramped up and then held at the maximum value, the capacitance starts rising as soon as the maximum value is reached. The inset shows the raw data, which displays the dipole gap behavior at long times. The dipole gap has been subtracted from the response in the main figure.

Although an exhaustive study was only performed on

BK7, the memory effect is present in a wide variety of materials. Five samples, AlBaSi (50 μm thick)[10], BK7 (150 μm and 300 μm) [11], Kapton (8 μm) [12], Mylar (15 μm) [13], and Coverglass (225 μm) [14], were studied, and all but Mylar have shown memory effects. Of the four materials which display the effect, one of them, Kapton, is a polymer, and another, AlBaSi, is a multicomponent glass.

To test for possible experimental heating, we cooled down two samples of BK7 which were identical except that one was 300μm thick and one was 150μm. Although the sweep rates and field sizes were the same for both samples, the size of the memory effect was larger in the thick sample. This might indicate that the heat generated within the sample contributes to the effect, since for the same excitation field twice as much heat is generated in the thick sample due to dissipation. Clearly, however, the effect cannot just be due to the sample changing temperature, since this would never explain the edge drops which we describe above. The sudden jump in capacitance upon reversing the sweep direction is also not consistent with a model which assumes that the memory effect is due to the sample changing temperature. There must be something about the sample during the first sweep through a given voltage that is different from all subsequent sweeps. A more plausible possibility is that some TLS start out in meta-stable non-equilibrium states, but that the sweep enhances their relaxation towards thermal equilibrium. Thus, performing the sweep for the first time enables some TLS to come into thermal equilibrium to a state with a higher dielectric constant. However, the memory effect decays away if we wait for a long time at zero field, suggesting that the sample is not in an equilibrium state. While we cannot definitively rule out a heating effect, the presence of the edge drop, the turnaround drop, and the decay of the memory effect all indicate that the effect is not simply a trivial one arising from heating of the sample.

The fact that various samples show a memory effect suggests that the tunneling systems are responsible for this effect. The memory effect depends on the sweep rate and the time between various sweeps, further indicating that relaxation is important. With the speed of sound in BK7 of about $c \simeq 5000$ m/s the wave length of a phonon with an energy corresponding to $E/k_B = 10$ mK is $\lambda \simeq 25$ μm. This is just a sixth of the thickness of the thin sample, and it is possible that surface effects start to influence the coupling to the TLS and thus their relaxation times. However, further experimental and theoretical investigations are necessary to resolve the issue.

One of us has proposed [15] that the memory effect is due to resonant pairs (RPs), weakly coupled pairs of tunneling systems fulfilling

$$|E_1 - E_2| \ll J \ll E_1 \simeq E_2 =: E \quad (1)$$

where $E_i = \sqrt{\Delta_i^2 + \Delta_{0i}^2}$ is the energy of TLS i and J

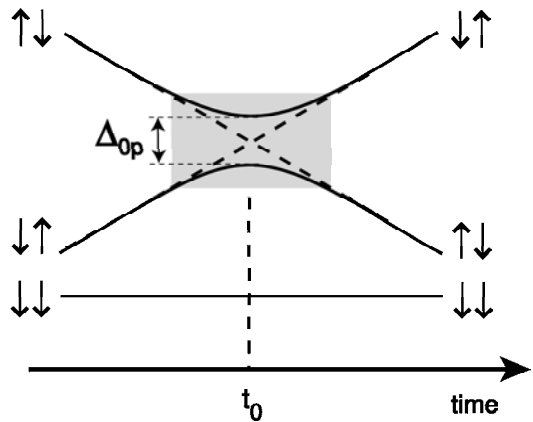

FIG. 4: Energy spectrum of a pair of TLS while an external electric field is swept across the sample. The grey region, which is magnified for clarity, indicates the "resonant region", where the two TLS form a resonant pair.

is the coupling energy between the tunneling systems. The coupling yields an avoided crossing between the two excited states of the single TLS, which results in an excess contribution to the dielectric constant. In thermal equilibrium this contribution is negligible since the occupation difference between the split states is vanishingly small. However, during a field sweep the TLS are far from equilibrium. In Fig. 4 the spectrum of a pair of TLS which form a RP in the shaded region is plotted over time during a field sweep. Outside the shaded region the interaction is negligible and the two TLS are effectively uncoupled. Using notation familiar from spin systems, we will denote a TLS in its excited state by ↑ and one in its ground state by ↓. A substantial occupation difference between the split states for a RP formed during a field sweep is expected for thermal TLS ($E \sim 2k_BT$) where one TLS has a relaxation time equal to the time the field needs to change the asymmetry energies by $\vec{p}\delta\vec{F}(\tau*) = 2k_BT$. Note that on average a RP formed by two TLS with relaxation times $\tau*$ has an occupation difference twice as large [15].

If a pair is initially in the state ↓↑ and we drive it adiabatically through the region where the RP is formed, it will evolve into the state ↑↓. Thus, the formation of a resonant pair while sweeping results in a flip-flop. Consider a TLS which forms a RP with a second TLS at $t_0$ and whose relaxation time is long compared to any experimental time scale. When the RP is formed this TLS is still in its original state. The flip-flop exchanges the occupation numbers between the two TLS so the state of the second TLS will be stored in the TLS with the very long relaxation time.

This explains the "turnaround jump" and the decay of the "edge drop", which simply results from the very slow relaxation of the second TLS towards equilibrium. The relaxation rate due to the one-phonon pro-cess is given by $\tau^{-1} = \gamma_0 \Delta_0^2 E \coth(\beta E/2)$ with $\gamma_0 = (1/c_l^5 + 2/c_t^5)B^2/(2\pi\rho\hbar^4)$, the longitudinal and transversal speed of sound $c_{l,t}$, the density of the glass $\rho$ and the strain-coupling constant $B$ [16]. Thus, the optimal relaxation time $\tau*$ corresponds to a tunneling splitting

$$\Delta_0^\star = \sqrt{\frac{\vec{p}\vec{F}}{t_R} \frac{1}{2\gamma_0 T^2}} \qquad (2)$$

Nalbach [15] obtains for the difference in dielectric constant between sweeping up in field for the first time and sweeping down:

$$\frac{\delta \epsilon'^\star}{\epsilon'^\star} = 1.6 \cdot P_0 p^2 \frac{k_B T}{E_{max}} \ln\left(\frac{\Delta_0^\star}{\Delta_{0min}}\right) \qquad (3)$$

where $p$ is the average dipole moment, $E_{max}$ is the high energy cutoff for $E$ and $\Delta_{0min}$ is the minimum tunneling element.

The drop seen as the sweep starts and the increase upon stopping a ramp are easily understood within the context of this model once we realize that, because the energy associated with the AC excitation is around 1 mK, the system is constantly performing many small sweeps. This is consistent with the width of the initial drop and turnaround jump being of order the temperature.

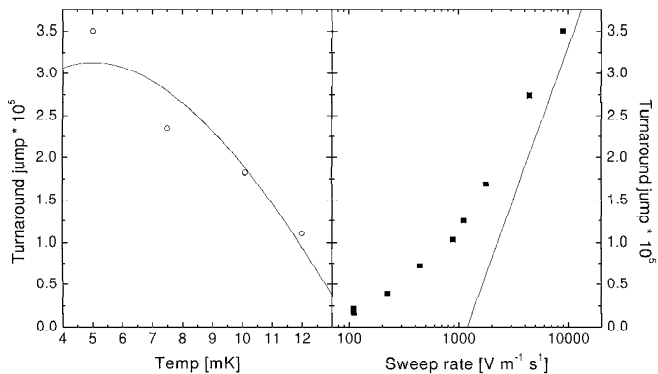

FIG. 5: The turnaround jump increases with decreasing temperature and increasing sweep rate. The data is in good agreement with a fit using Nalbach's theory.

Fig. 5 shows data and fits from Nalbach's theory, assuming that one-phonon thermal relaxation dominates. The parameters were first extracted from the temperature dependence, and these parameters were then used to create the fit to the sweep rate. We obtain reasonable fits to the data using $\Delta_{0min} = 240\mu K$, $E_{max} = 1K$, $P_0 p^2 = 3.9 \cdot 10^{-3}$, and $\gamma_0 = 1 \cdot 10^8 K^{-3} s^{-1}$. The overall size of the effect depends on $P_0 p^2/E_{max}$. Using the common estimate $E_{max} = 1\ K$, we find that our fit value $P_0 p^2 = 3.9 \cdot 10^{-3}$ agrees well with previous measurements [17]. Both the fit to the temperature variation and to the sweep rate dependence use the same set of parameters. The theory neglects driving by the AC field. With decreasing sweep rate the AC field driving gains importance and the data for low sweep rates might be strongly influenced by the AC driving, while for the fast sweep rates the DC driving dominates.

We have seen striking new behavior in the low temperature response of dielectric solids to large DC fields. The theory provides a qualitative description of the data. Exploring this effect further could aid in the understanding of the importance of interactions in determining the low temperature behavior of amorphous solids.

The authors wish to thank A. Burin for helpful discussions.